# Atomically thin mica flakes and their application as ultrathin insulating substrates for graphene

*Andres Castellanos-Gomez[1,2,+,*], Magdalena Wojtaszek[2], Nikolaos Tombros[2,3], Nicolás Agraït[1,4,5], Bart J. van Wees[2] and Gabino Rubio-Bollinger[1,4,*].*

[1] Departamento de Física de la Materia Condensada. Universidad Autónoma de Madrid, Campus de Cantoblanco, E-28049 Madrid, Spain.
[2] Physics of Nanodevices, Zernike Institute for Advanced Materials, University of Groningen, The Netherlands.
[3] Molecular Electronics, Zernike Institute for Advanced Materials, University of Groningen, The Netherlands.
[4] Instituto Universitario de Ciencia de Materiales "Nicolás Cabrera". Universidad Autónoma de Madrid, Campus de Cantoblanco, E-28049 Madrid, Spain.
[5] Instituto Madrileño de Estudios Avanzados en Nanociencia. IMDEA-Nanociencia, E-28049 Madrid, Spain.
[+] Present address: Kavli Institute of Nanoscience, Delft University of Technology, Lorentzweg 1, 2628 CJ Delft (The Netherlands)

[*] E-mail: a.castellanosgomez@tudelft.nl , gabino.rubio@uam.es

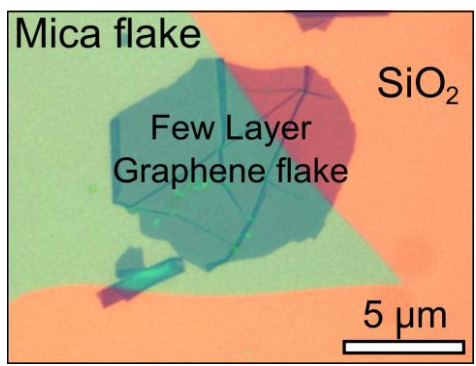

*Optical micrograph of a 3 nm thick few layer graphene flake transferred, with an all-dry transfer technique based on the use of viscoelastic stamps, on top of a 12 nm thick mica flake, mechanically exfoliated on top of a $SiO_2$/Si surface.*

We show that it is possible to deposit, by mechanical exfoliation on $SiO_2$/Si wafers, atomically thin mica flakes down to a single monolayer thickness. The optical contrast of these mica flakes on top of a $SiO_2$/Si substrate, which depends on their thickness, the illumination wavelength and the $SiO_2$ substrate thickness, can be quantitatively accounted for by a Fresnel law based model. The preparation of atomically thin insulating crystalline sheets will enable the fabrication of ultrathin defect-free insulating substrates, dielectric barriers or planar electron tunneling junctions. Additionally, we show that few-layer graphene flakes can be deposited on top of a previously transferred mica flake. Our transfer method relies on viscoelastic stamps, as those used for soft lithography. A Raman spectroscopy study shows that such an all-dry deposition technique yields cleaner and higher quality flakes than conventional wet-transfer procedures based on lithographic resists.

**1. Introduction**

The experimental realization of graphene, just a single atomic layer of graphite, by mechanical exfoliation of graphite on $SiO_2$ [1, 2] surfaces has paved the way to study a very interesting family of two-dimensional crystals almost unexplored so far. Apart from graphene, mechanical exfoliation has been used to prepare other atomically thin crystals [1] such as $MoS_2$,[3-6] a semiconductor. However, apart from conducting and semiconducting materials, the microelectronic industry also needs insulators which can be used as substrates, dielectrics or electron tunnelling barriers. Mechanical exfoliation also enables the production of insulating atomically thin crystals but up to now the fabrication has been restricted to a few insulating two dimensional crystals such as hexagonal boron nitride.[7, 8]

The layered structure of muscovite mica, a phyllosilicate mineral of aluminum and potassium with chemical formula $(KF)_2(Al_2O_3)_3(SiO_2)_6(H_2O)$, makes this material a promising candidate to produce atomically thin insulating crystals by mechanical exfoliation. Moreover, due to its high resistance to heat, water and chemical agents, to its mechanical properties and to its high dielectric constant, bulk mica has been already extensively used in the electronic industry in many applications as substrate, heat and electrical insulator or dielectric barrier. Recently, bulk mica has been also employed as a substrate to fabricate ultraflat graphene samples.[9] It has been observed that graphene deposited on mica adheres to the atomically flat terraces of mica without noticeable rippling. Note that the ripples in graphene, unavoidable when it is either deposited on $SiO_2$ [10] or suspended,[11] can modify its electronic properties and induce charge inhomogeneities.[12, 13] It would be therefore very interesting to study how the atomically flat topography of graphene on mica affects the electronic properties of graphene. The use of a bulk mica substrate for graphene-based devices, however, hampers the capability of electrostatic doping of graphene with a backgate. This limitation was overcome by employing 10-50 nm thick mica flakes deposited on $SiO_2$/Si substrates.[14] The two dimensional nature of these atomically thin crystalline insulator sheets makes them very interesting candidates in applications such as insulating barriers in planar tunnel junctions or as flexible substrates for graphene or molecular electronics devices. However, despite the large number of potential applications of these crystalline insulator nanosheets, the details about the fabrication, identification and characterization of these ultrathin mica flakes are missing in the literature.

In this work we report the fabrication of atomically thin mica flakes as thin as just one layer thick (1 nm) on $SiO_2$/Si substrates. We also present a combined characterization by quantitative optical microscopy and atomic force microscopy of these flakes. From this study one can determine the optimal $SiO_2$ substrate thickness and illumination wavelength to reliably identify atomically thin crystals of mica by optical microscopy. In addition, we show an all-dry procedure to transfer few-layers graphene (FLG) flakes on top of atomically thin mica flakes, based on viscoelastic stamps like the ones used in soft-lithography. From Raman spectroscopy measurements, we conclude that this transfer technique produces cleaner and higher quality flakes than conventional wet-transfer procedures based on lithographic resist.[8, 15]

## 2. Results

### *2.1. Sample fabrication*

In order to produce atomically thin muscovite mica flakes, we have employed the micromechanical cleavage technique, widely known from fabrication of graphene flakes [2] and other materials.[16] To cleave the starting material this technique usually employs an adhesive tape which can leave traces of glue on the surface that contaminate the fabricated sample.[17] This problem can be avoided by replacing the adhesive tape by a viscoelastic stamp, similar to the ones used in soft-lithography.[18] In previous works, we successfully used stamps of poly (dimethil)-siloxane (PDMS) to fabricate graphene,[19] $NbSe_2$ and $MoS_2$ [3] atomically thin crystals. To produce atomically thin mica flakes we start cleaving a bulk muscovite mica sample by pressing the surface of the stamp against the bulk mica and peeling it off suddenly. After this step of the process part of the mica surface is cleaved and transferred to the stamp surface. Next, to transfer the mica flakes to an arbitrary surface one presses the surface of the stamp against the receptor surface and peels it off slowly (5 seconds). Using this technique we have produced mica flakes with thicknesses ranging from hundreds of nanometers down to just one nanometer, which is the thickness of a single layer of mica.

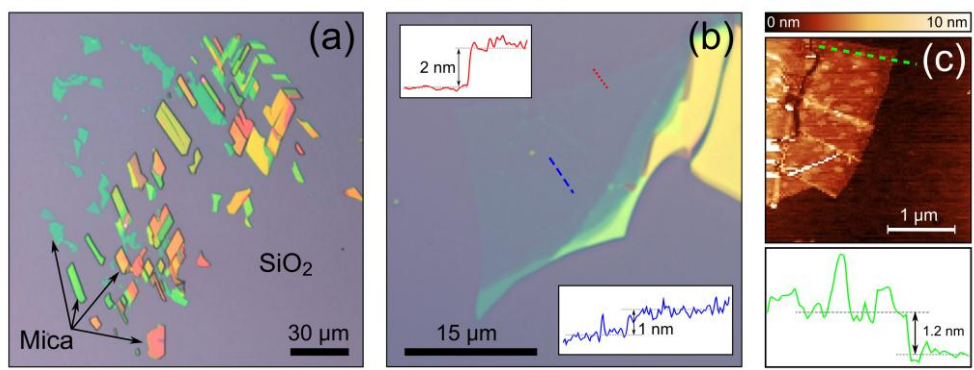

**Figure 1.** (a) Optical micrograph of several mica flakes deposited on a 300 nm $SiO_2$/Si substrate by mechanical exfoliation. The different colors of the flakes correspond to different thicknesses. (b) Optical micrograph of an ultrathin muscovite mica flake on a 300 nm $SiO_2$/Si substrate. (Top insert in b) Contact mode AFM topographic line profile measured across the interface between the mica flake and the $SiO_2$ substrate (dotted red line). The thickness of the flake in this region is 2 nm which corresponds to the thickness of two single mica layers. (Bottom insert in b) The topographic profile (along the dashed blue line) showing the boundary between a region 2 nm thick and another 3 nm thick. (c) Contact mode AFM topographic image of a mica flake with a thickness of 1.2 nm, corresponding to a single layer of mica. A topographic profile along the green dashed line is shown below panel (c).

**Figure 1(a)** shows some mica flakes transferred onto a silicon substrate with a 300 nm SiO$_2$ capping layer. We have observed that the apparent color of a flake depends on its thickness. Figure 1(b) presents an optical micrograph under white illumination of a mica flake, with thickness ranging from 2 nm to 20 nm. The top insert in Figure 1(b) shows a contact mode atomic force microscopy (AFM) topographic profile measured through the boundary between the SiO$_2$ surface and the mica flake. The height difference between the SiO$_2$ substrate and the mica flake in this region is 2 nm, which corresponds to the thickness of two single layers of muscovite mica. The bottom insert in Figure 1(b) shows the contact mode AFM topographic profile measured across the boundary between a region two layers thick and other three layers thick (3 nm thickness). Although we find that single layer flakes are usually not much larger than $1 \times 1$ μm$^2$ (Figure 1(c)), the typical area of thin mica flakes (2 or 3 layers) can be up to $5 \times 5$ μm$^2$. The large area of these nanometre thick mica sheets makes them promising candidates for their use not only as substrate for graphene-based devices but also as insulator barriers in planar tunnel junctions or dielectrics in nanocapacitors.

*2.2. Optical characterization*

The physics behind the optical visibility of these atomically thin crystals can be illustrated with the example of the interference colors in SiO$_2$ thin films. It is well known that the thickness of thermally grown SiO$_2$ layers on Si wafers can be determined with ~ 5 nm accuracy readily from their color under white light illumination.[20, 21] This apparent color is due to the interference of the paths reflected at the air/SiO$_2$ and SiO$_2$/Si interfaces (similar to a Fabry-Perot interferometer). The interfering paths will have a relative phase shift which depends on the illumination wavelength and the SiO$_2$ layer thickness, hence the SiO$_2$ thickness dependence of the color of the wafers. If one deposits a thin mica sheet instead of a SiO$_2$ layer the effect should be similar. If the mica sheet is atomically thin the optical path difference could be small and the color difference between the mica sheet and the bare Si wafer could thus be imperceptible. The optical visibility of atomically thin crystals, however, can be enhanced by depositing them on a multilayered medium, typically a Si wafer with a thermally grown SiO$_2$ capping layer.[22]

To interpret the observed optical contrast one can employ a simple model based on the Fresnel law.[22, 23] This approach has been successfully employed to study the optical contrast of different two dimensional crystals such as graphene[22-24], transition metal dichalcogenides MoS$_2$ and NbSe$_2$[3, 25] or hexagonal boron nitride.[26] The subscripts 0, 1, 2 and 3 will label hereafter the air, mica flake, SiO$_2$ and Si media respectively. We consider normal incidence of the light through the trilayer structure formed by the mica flake, the SiO$_2$ and the Si. The optical properties of the Si layer, considered semi-infinite, are given by its complex refractive index $\tilde{n}_3(\lambda) = n_3 - i\kappa_3$ which strongly depends on the illumination wavelength ($\lambda$) in the visible range of the spectrum.[27] The SiO$_2$ layer, with a thickness $d_2$, is described by its refractive index $\tilde{n}_2(\lambda)$ which also depends on the illumination wavelength.[27] Note that using the refractive indexes of $SiO_2$ and Si one can accurately account for the interference colors of the oxidized wafers with a Fresnel law based model.[20, 21] The reflected intensity for a $SiO_2$/Si wafer ($I_0$) can be expressed in terms of the phase shift produced by the $SiO_2$ layer ($\Phi_2 = 2\pi \tilde{n}_2 d_2/\lambda$) and the amplitudes of the paths reflected at the air/$SiO_2$ and $SiO_2$/Si interfaces ($r_{02}$ and $r_{23}$ respectively),

$$I_0(\lambda) = \left| \frac{r_{02} + r_{23} e^{-2i\Phi_2}}{1 + r_{02} r_{23} e^{-2i\Phi_2}} \right|^2 \quad (1)$$

where the amplitude of the reflected path in the interface between the media $i$ and $j$ is $r_{ij} = (\tilde{n}_i - \tilde{n}_j)/(\tilde{n}_i + \tilde{n}_j)$ with $\tilde{n}_j$ being the refractive index of medium $j$. The atomically thin mica crystal is taken into account as a layer of thickness $d_1$ on the $SiO_2$ medium whose refractive index is $\tilde{n}_1(\lambda) = n_1 - i\kappa_1$. The reflected intensity from the mica flake ($I_0$) can be written as,[22, 23]

$$I_1(\lambda) = \left| \frac{r_{01} e^{i(\Phi_1+\Phi_2)} + r_{12} e^{-i(\Phi_1-\Phi_2)} + r_{23} e^{-i(\Phi_1+\Phi_2)} + r_{01} r_{12} r_{23} e^{i(\Phi_1-\Phi_2)}}{e^{i(\Phi_1+\Phi_2)} + r_{01} r_{12} e^{-i(\Phi_1-\Phi_2)} + r_{01} r_{23} e^{-i(\Phi_1+\Phi_2)} + r_{12} r_{23} e^{i(\Phi_1-\Phi_2)}} \right|^2 \quad (2)$$

$\Phi_1 = 2\pi \tilde{n}_1 d_1/\lambda$ is the phase shift of the path produced by the presence of the mica flake. Using the expressions (1) and (2) one can obtain the optical contrast ($C$) which is defined as

$$C = \frac{I_1 - I_0}{I_1 + I_0} \quad (3)$$

**Figure 2(a)** shows the optical contrast, measured at different illumination wavelengths, for flakes 2 to 10 layers thick. The solid lines in the plot are the contrast *vs*. wavelength dependencies according to expressions (1), (2) and (3). The thickness of the flake has been determined by contact mode AFM and the refractive index of the mica flake is taken to be $\tilde{n}_1 = 1.55 - 0i$, that is independent from the illumination wavelength and similar to the one of bulk muscovite mica. The grayed areas enveloping the solid lines mark the calculated contrast *vs*. wavelength dependencies considering an uncertainty of the real part of the refractive index $\Delta n_1 = \pm 0.1$. The measured optical contrast values are in good agreement with the calculated data indicating, within experimental resolution, that the refractive index of the atomically thin mica flakes can be considered close to its bulk value. Notice that in Figure 2(a) the experimental contrast is systematically slightly lower than the calculated one. This was observed in similar experiments [23] and was attributed to the finite numerical aperture of the microscope objective.[25] Figure 2(b) shows the optical contrast maps for the same mica flake shown in Figure 1(b) at 500 nm, 546 nm, 568 nm and 632 nm illumination wavelengths. In agreement with the data shown in Figure 2(a), at $\lambda$ = 568 nm the optical contrast of the thinnest part of the flake vanishes while it shows a maximum (minimum) at $\lambda$ = 520 nm (630 nm).

- 5 -

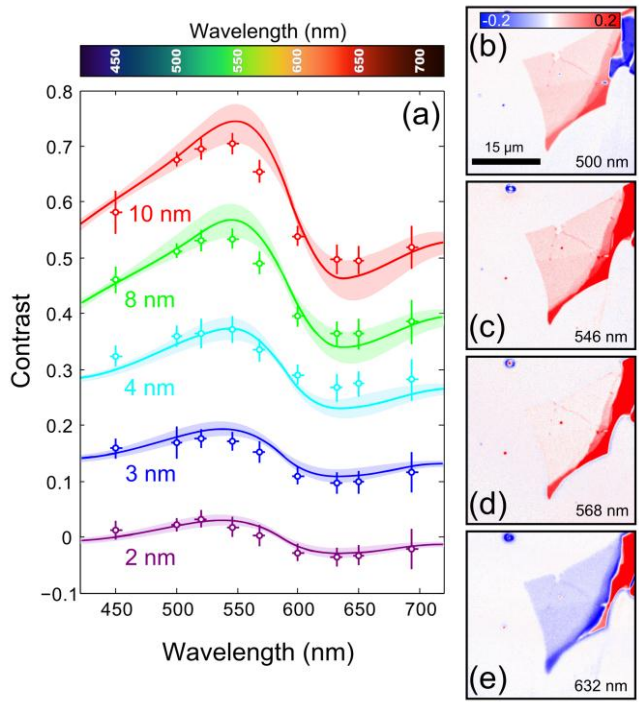

**Figure 2:** (a) Optical contrast measured at different illumination wavelengths on mica flakes from 2 to 10 nm thick. The contrast *vs*. wavelength dependence calculated with the Fresnel law based model is also shown (solid lines). The refractive index of mica flakes is $n = 1.55$ ($\kappa = 0$) and an uncertainty in its value $\Delta n \pm 0.1$ has been considered and displayed as the grayed region enveloping the solid lines. Note that the data for the 3 nm, 4 nm, 8 nm and 10 nm have been vertically displaced for clarity by 0.15, 0.3, 0.45 and 0.6. (b-e) Optical contrast maps measured, in the mica flake shown in Figure 1, at different illumination wavelengths.

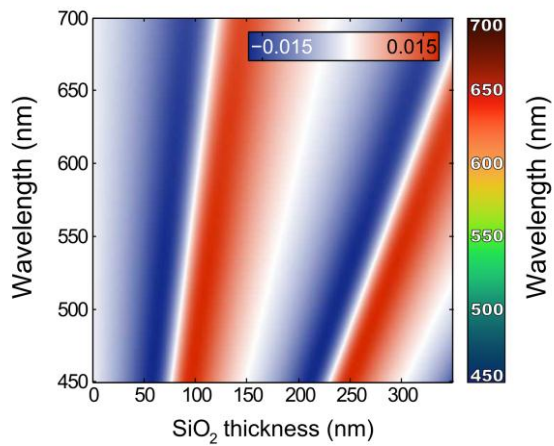

**Figure 3:** Color map of the calculated optical contrast for a monolayer mica flake as a function of the illumination wavelength and the $SiO_2$ thickness. A color bar is inserted in the plot.

Expressions (1), (2) and (3), can also be used to calculate the optical contrast yielded by a single layer of mica for different illumination wavelengths and $SiO_2$ capping layer thicknesses. **Figure 3** shows the result of this calculation which can be used as a guide to find the appropriate conditions to identify ultrathin mica flakes atop $SiO_2$/Si substrates. We have selected the $SiO_2$ thicknesses which optimize the contrast for $\lambda = 550$ nm that is the illumination wavelength to which the human eye attains maximum sensitivity.[29] The first four different $SiO_2$ thicknesses which optimize the optical contrast for a single layer of mica are: 55 nm (-1.5% contrast, nearly $\lambda$ independent), 100 nm (+1.5% contrast, also nearly $\lambda$ independent), 260 nm (-1.5% at $\lambda = 550$ nm and 0% at $\lambda = 500$ nm) and 305 nm (+1.5% at $\lambda = 550$ nm and 0% at $\lambda = 580$ nm). Note that 1.5% contrast is roughly the detection limit of the human eye and thus special care has to be taken to optimize both illumination wavelength and SiO$_2$ thickness in order to identify these ultrathin mica flakes by eye.

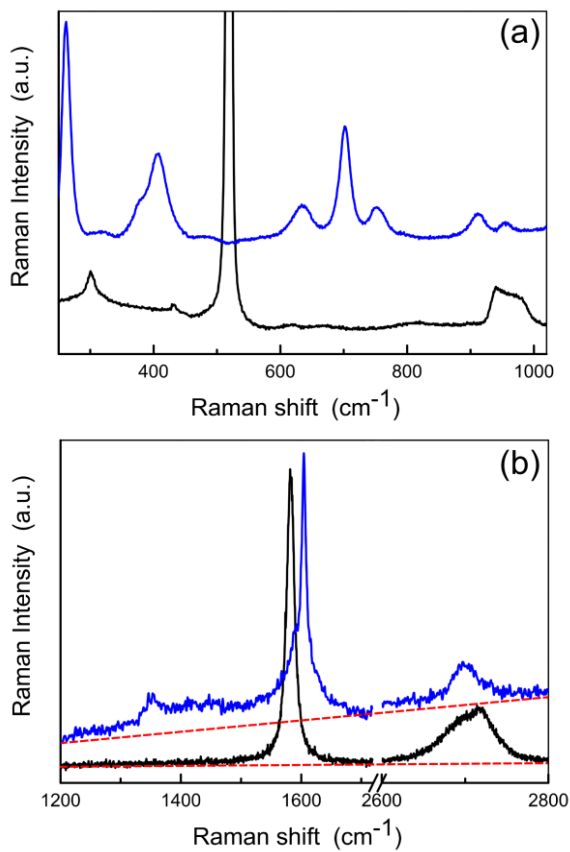

**Figure 4:** (a) Raman spectra of a thin mica flake (less than 12 nm thick) on SiO$_2$/Si (black) and bulk mica (blue) – a parent material for analyzed flakes. For mica flakes from 2 nm to 60 nm thick Raman spectra showed only features of the Si underneath. No peaks corresponding to mica vibrational modes can be resolved. (b) Raman spectrum of multi layer graphene after PDMS-based transfer (black) and the spectrum of single layer graphene after PMMA-based transfer (blue). No baseline corrections were applied to the spectra. The strong linear increase of the background, emphasized by a red dashed line as an eye-guide, for the case of PMMA-based transfer is apparent.

It is a common practice to assist the optical characterization of layered materials with their Raman spectra. Such spectra turned out to be especially informative in the case of graphene, where the precise distinction between single, double and multilayer samples is possible.[30, 31] We measured the Raman spectrum for the mica flakes of different thicknesses, ranging from 2 nm (2 layers) to 10 nm (10 layers) deposited on 300 nm SiO$_2$ and for bulk mica. None of the spectra of thin mica flakes showed typical Raman bands of bulk mica, even after long acquisition times (5 min.), presenting only the spectral features of its substrate – Silicon (a strong Lorentzian peak at ~520 cm$^{-1}$ and a broad flat band between 930 and 1000 cm$^{-1}$) – see **Figure 4(a)**. For comparison, we measured also Raman spectra of the parent material, bulk mica, containing all typical Raman bands known from the literature (Figure 5(b)).[32, 33] From this we conclude that the thin mica flakes give a too weak signal to be detected in Raman, excluding Raman spectroscopy as a useful tool for its detection and investigation. Invisibility of the mica flakes in Raman spectra carries however a large advantage because it means that spectra won't interfere with the Raman signature of other material, when deposited on top. Such an interference of Raman signatures occurs in the case of graphene, when deposited on Hexagonal Boron Nitride (BN), which was proposed as an advantageous new substrate for graphene electronics due to its flat and free from trapped charges surface.[8] In BN a main vibrational Raman mode (E2g) occurs at ~1366cm$^{-1}$ and its large intensity can screen in the spectrum the D-band of graphene (which occurs at 1344cm$^{-1}$ and as induced by defects gives a valuable information about the flake quality). From this point of view the invisibility of the thin mica flakes in Raman ensures the proper detection of all graphene Raman bands when deposited on top.

*2.3. Transfer of FLG on top of thin mica flakes*

As pointed out in the introduction, ultrathin mica flakes can be very appealing substrates to fabricate graphene electronic devices. By depositing graphene on top of an ultrathin mica flake one can uncouple the graphene flake from the SiO$_2$ substrate while maintaining the possibility of applying an electric field through the SiO$_2$/mica to change the graphene doping. In this work we present an all-dry procedure to transfer few-layers graphene (FLG) flakes on top of thin mica flakes.

The main steps of the procedure are depicted in **Figure 5(a)**. First, a PDMS stamp is used to cleave an HOPG graphite sample. Second, FLG flakes are identified on the PDMS surface by optical inspection in transmission mode optical microscopy. Third, because the PDMS stamp is transparent, the selected FLG flake can be accurately aligned to the target mica flake, using a 3-axis micropositioner stage attached to an optical microscope, which has been previously deposited on a 300 nm SiO$_2$/Si substrate. Forth, the PDMS stamp is brought into contact with the substrate and subsequently peeled off very slowly (5 seconds). In this way the FLG flake is transferred on top of the mica flake. Figure 5(b) shows an optical microscopy image of a 10 layers thick graphene flake deposited on top of a 12 layers thick mica flake. Additionally, we inspect the graphene flake using Raman spectroscopy. As Raman measurements are sensitive to fluorescent background, they give information not only about the quality of the transferred flake but also about the organic/polymer traces left after the transfer process. In Figure 4 we compare the Raman spectra of the flake transferred to SiO$_2$ with PDMS stamps and the flake transferred with a wet-transfer technique based on PMMA resist.[8, 15] The acquisition conditions were the same for both samples, with the same setup used to study the mica flake. The flakes transferred with PDMS stamps only show a small constant background, which can be attributed to thermal fluctuations of the CCD detector. The flake also does not exhibit defects (no D–band at 1344cm$^{-1}$) and well resemble spectra of the freshly cleaved, pristine few layer graphene. On the contrary, the flake obtained by transferring with PMMA shows small D-band (5% of the G-band intensity), indicating some lattice defects or adsorbates and a much larger linearly increasing background, which is a signature of fluorescent contaminants. The ratio between G-band (~1580cm$^{-1}$) and 2D-band (~2680cm$^{-1}$) is below 1, indicating high doping of the flake. We have additionally found that all the Raman signatures of the few layers graphene flake transferred on top of the thin mica flake remains unmodified proving the convenience of ultrathin mica flakes in applications that require atomically flat substrates yet inactive to Raman spectroscopy. From that one can conclude that this all-dry transfer technique yields flakes of much higher quality, which are desirable for making the most of its remarkable electronic properties.

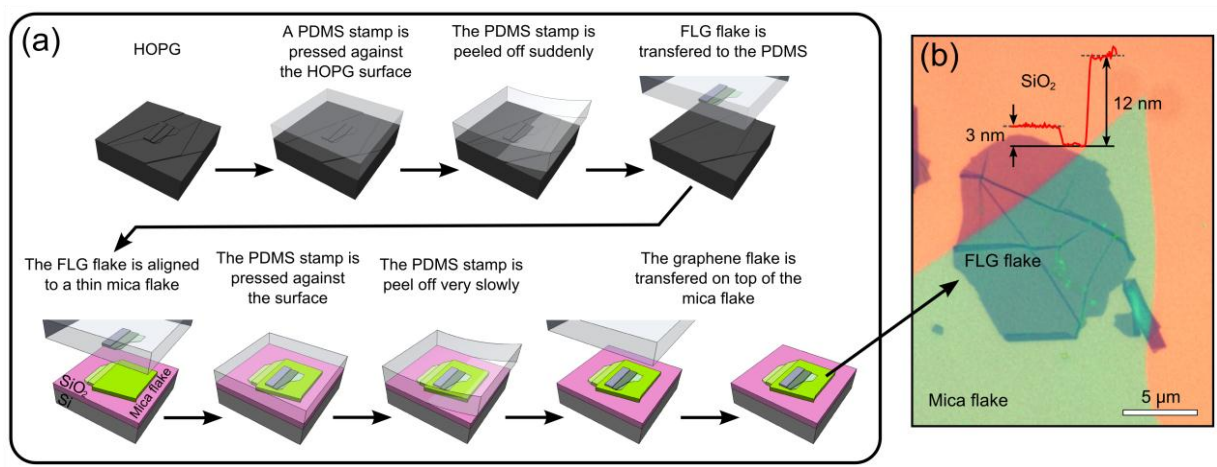

**Figure 5:** (a) Schematic diagram of the transfer procedure to deposit FLG flakes on top of a thin mica flake. We first cleave a bulk HOPG sample with a PDMS stamp. Then the stamp is inspected with an optical microscope in transmission mode to find thin graphite flakes. The selected flake is aligned to a thin mica flake previously deposited on a 300 nm $SiO_2$/Si substrate by micromechanical cleavage. The stamp is brought into contact with the surface and it is peeled off very slowly. The FLG flake is then transferred on top of the mica flake. (b) Optical micrograph of a 3 nm thick FLG on top of a 12 nm thick mica flake. The AFM topographic profile measured along the solid black line is also shown in the image.

## 3. Conclusions

We have deposited atomically thin mica flakes on $SiO_2$/Si wafers using a procedure based on mechanical exfoliation with viscoelastic PDMS stamps. This procedure yields mica crystals whose thickness can be as small as that of a single layer. We have studied the optical contrast of flakes with thicknesses ranging from 2 nm to 10 nm, measured at different illumination wavelengths. We find that the measured optical contrast can be accurately accounted for by a model based on the Fresnel law. Using this model we have determined the optimal silicon oxide thickness and illumination wavelength to reliably identify, by optical microscopy, atomically thin mica flakes deposited on $SiO_2$/Si substrates. In addition, we have developed an all-dry procedure to transfer few-layers graphene flakes on top of atomically thin mica flakes, using again viscoelastic stamps. Raman spectra measurements have been used to characterize the quality of the transferred flake and also to get information about the organic/polymer traces left after the transfer process. We have found that our transfer technique produces cleaner and higher quality flakes than conventional wet-transfer procedures based on lithographic resist such as PMMA. The experimental realization of these crystalline atomically thin insulating sheets expands an interesting family of two-dimensional crystals. Moreover, the combination of these mica sheets with other materials such as graphene or $MoS_2$ can be used to engineer atomically thin crystalline heterostructures.

### 4. Experimental Section

The poly (dimethil)-siloxane (PDMS) stamps, used during the mica flake preparation and the all-dry transfer procedure, have been cast by curing the Sylgard® 184 elastomer kit purchased from Dow Corning.[3]

The atomic force microscope (AFM) used for the thickness measurement of the flakes is a Nanotec Cervantes AFM (Nanotec Electronica) operated in contact mode under ambient conditions. We have selected contact mode AFM instead of dynamic modes of operation to avoid artifacts in the determination of the flakes thickness.[34]

The quantitative measurements of the optical contrast of ultrathin mica flakes has been done with a Nikon Eclipse LV100 optical microscope under normal incidence with a 50× objective (numerical aperture NA = 0.8) and with a digital camera EO-1918C 1/1.8" (from Edmund Optics) attached to the microscope trinocular. The illumination wavelength was selected by means of nine narrow band-pass filters (10 nm FWHM) with central wavelengths 450 nm, 500 nm, 520 nm, 546 nm, 568 nm, 600 nm, 632 nm, 650 nm y 694 nm purchased from Edmund Optics.

The Raman spectra were recorded at room temperature using Horiba T64000 Raman spectrometer with 532 nm laser excitation wavelength of unpolarized light (resolution ~2 $cm^{-1}$). The spot size was <10 μm in diameter and the power density was about 30 $W/cm^2$.

### Acknowledgements

A.C-G. acknowledges fellowship support from the Comunidad de Madrid (Spain). This work was supported by MICINN (Spain) (MAT2008-01735 and CONSOLIDER en Nanociencia molecular CSD-2007-00010). M.W. is grateful to Optical Condensed Matter Group at University of Groningen (Netherlands) and especially to Ben Hesp for the access to Raman facilities and help with measurements. M.W. and N.T. acknowledge financial support from Ubbo Emmius program of Groningen Graduate School of Science and Zernike Institute for Advanced Materials, respectively.

### References

[1]     K. Novoselov, D. Jiang, F. Schedin, T. Booth, V. Khotkevich, S. Morozov, A. Geim, P. Natl. Acad. Sci. USA 2005, 102, 10451.
[2]     K. S. Novoselov, A. K. Geim, S. V. Morozov, D. Jiang, Y. Zhang, S. V. Dubonos, I. V. Grigorieva, A. A. Firsov, Science 2004, 306, 666.
[3]     A. Castellanos-Gomez, N. Agrait, G. Rubio-Bollinger, Appl. Phys. Lett. 2010, 96, 213116.
[4]     C. Lee, H. Yan, L. E. Brus, T. F. Heinz, J. Hone, S. Ryu, ACS nano 2010, 4, 2695.

[5]     H. S. S. Ramakrishna Matte, A. Gomathi, A. K. Manna, D. J. Late, R. Datta, S. K. Pati, C. N. R. Rao, Angew. Chem. 2010, 122, 4153.

[6]     A. Splendiani, L. Sun, Y. Zhang, T. Li, J. Kim, C.-Y. Chim, G. Galli, F. Wang, Nano Lett. 2010.

[7]     D. Pacile, J. Meyer, Ç. Girit, A. Zettl, Appl. Phys. Lett. 2008, 92, 133107.

[8]     C. Dean, A. Young, I. Meric, C. Lee, L. Wang, S. Sorgenfrei, K. Watanabe, T. Taniguchi, P. Kim, K. Shepard, Nature Nanotech. 2010, 5, 722.

[9]     C. Lui, L. Liu, K. Mak, G. Flynn, T. Heinz, Nature 2009, 462, 339.

[10]    V. Geringer, M. Liebmann, T. Echtermeyer, S. Runte, M. Schmidt, R. Rückamp, M. C. Lemme, M. Morgenstern, Phys. Rev. Lett. 2009, 102, 076102/1.

[11]    J. C. Meyer, A. K. Geim, M. I. Katsnelson, K. S. Novoselov, T. J. Booth, S. Roth, Nature 2007, 446, 60.

[12]    M. L. Teague, A. P. Lai, J. Velasco, C. R. Hughes, A. D. Beyer, M. W. Bockrath, C. N. Lau, N. C. Yeh, Nano Lett. 2009, 9, 2542.

[13]    A. Castellanos-Gomez, R. H. Smit, N. Agrait, G. Rubio-Bollinger, Submitted 2010.

[14]    L. Ponomarenko, R. Yang, T. Mohiuddin, M. Katsnelson, K. Novoselov, S. Morozov, A. Zhukov, F. Schedin, E. Hill, A. Geim, Phys. Rev. Lett. 2009, 102, 206603.

[15]    A. Reina, X. Jia, J. Ho, D. Nezich, H. Son, V. Bulovic, M. Dresselhaus, J. Kong, Nano Lett. 2008, 9, 30.

[16]    K. Kalantar-zadeh, J. Tang, M. Wang, K. L. Wang, A. Shailos, K. Galatsis, R. Kojima, V. Strong, A. Lech, W. Wlodarski, R. B. Kaner, Nanoscale 2010, 2, 429.

[17]    J. Moser, A. Verdaguer, D. Jiménez, A. Barreiro, A. Bachtold, Appl. Phys. Lett. 2009, 92, 123507.

[18]    M. A. Meitl, Z. T. Zhu, V. Kumar, K. J. Lee, X. Feng, Y. Y. Huang, I. Adesida, R. G. Nuzzo, J. A. Rogers, Nature Mater. 2006, 5, 33.

[19]    M. Moreno-Moreno, A. Castellanos-Gomez, G. Rubio-Bollinger, J. Gomez-Herrero, N. Agraït, Small 2009, 5, 924.

[20]    J. Kvavle, C. Bell, J. Henrie, S. Schultz, A. Hawkins, Opt. Express 2004, 12, 5789.

[21]    J. Henrie, S. Kellis, S. Schultz, A. Hawkins, Opt. Express 2004, 12, 1464.

[22]    S. Roddaro, P. Pingue, V. Piazza, V. Pellegrini, F. Beltram, Nano Lett. 2007, 7, 2707.

[23]    P. Blake, E. W. Hill, A. H. C. Neto, K. S. Novoselov, D. Jiang, R. Yang, T. J. Booth, A. K. Geim, Appl. Phys. Lett. 2007, 91, 063124.

[24]    I. Jung, M. Pelton, R. Piner, D. Dikin, S. Stankovich, S. Watcharotone, M. Hausner, R. Ruoff, Nano Lett. 2007, 7, 3569.

[25]    M. Benameur, B. Radisavljevic, J. Héron, S. Sahoo, H. Berger, A. Kis, Nanotechnology 2011, 22, 125706.

[26]    R. V. Gorbachev, I. Riaz, R. R. Nair, R. Jalil, L. Britnell, B. D. Belle, E. W. Hill, K. S. Novoselov, K. Watanabe, T. Taniguchi, Small 2011.

[27]    C. Herzinger, B. Johs, W. McGahan, J. Woollam, W. Paulson, J. Appl. Phys. 1998, 83, 3323.

[28]    C. Casiraghi, A. Hartschuh, E. Lidorikis, H. Qian, H. Harutyunyan, T. Gokus, K. Novoselov, A. Ferrari, Nano Lett. 2007, 7, 2711.

[29]    G. Wald, Science 1945, 101, 653.

[30]    D. Graf, F. Molitor, K. Ensslin, C. Stampfer, A. Jungen, C. Hierold, L. Wirtz, Nano Lett 2007, 7, 238.

[31]    D. Graf, F. Molitor, K. Ensslin, C. Stampfer, A. Jungen, C. Hierold, L. Wirtz, Solid State Commun. 2007, 143, 44.

[32]    D. McKeown, M. Bell, E. Etz, Am. Mineral. 1999, 84, 1041.

[33]    D. McKeown, M. Bell, E. Etz, Am. Mineral. 1999, 84, 970.

[34]    P. Nemes-Incze, Z. Osváth, K. Kamarás, L. P. Biró, Carbon 2008, 46, 1435.